\documentclass[aps,pra,preprint,superscriptaddress,longbibliography]{revtex4-2}

\usepackage{amsmath}
\usepackage{amsthm}
\usepackage{amsfonts}
\usepackage{amssymb}
\usepackage{xcolor,graphicx}
\usepackage{tikz}
\usepackage{color}
\usepackage[pdftex]{hyperref} 
\usepackage{longtable}
\usepackage{array}
\usepackage{dsfont}

\begin{document}
	
	\title{Non-classical light state transfer in $su(2)$ resonator networks}
	
	\author{A. F. Mu\~noz Espinosa }
	\email[e-mail: ]{a00512932@exatec.tec.mx}
	\affiliation{Tecnol\'ogico de Monterrey, Escuela de Ingenier\'ia y Ciencias, Ave. Eugenio Garza Sada 2501, Monterrey, N.L., Mexico, 64849}
	

	\author{R.-K. Lee}
	\email[e-mail: ]{rklee@ee.nthu.edu.tw}
	\affiliation{Institute of Photonics Technologies, National Tsing Hua University, Hsinchu 30013, Taiwan.}
	\affiliation{Department of Physics, National Tsing Hua University, Hsinchu 30013, Taiwan.}
	\affiliation{Physics Division, National Center for Theoretical Sciences, Taipei 10617, Taiwan.}
	\affiliation{Center for Quantum Technologies, Hsinchu 30013, Taiwan.}
	
	\author{B. M. Rodr\'iguez-Lara}
	\email[e-mail: ]{bmlara@tec.mx}
	\affiliation{Tecnol\'ogico de Monterrey, Escuela de Ingenier\'ia y Ciencias, Ave. Eugenio Garza Sada 2501, Monterrey, N.L., Mexico, 64849 }	
	
	\date{\today}
	
	\begin{abstract}
	    We use a normal mode approach to show full and partial state transfer in a class of coupled resonator networks with underlying $su(2)$ symmetry that includes the so-called $J_{x}$ photonic lattice.
        Our approach defines an auxiliary Hermitian coupling matrix describing the network that yields the normal modes of the system and its time evolution in terms of orthogonal polynomials. 	   
        These results provide insight on the full quantum state reconstruction time in a general $su(2)$ network of any size and the full quantum transfer time in the $J_{x}$ network of size $4 n + 1$ with $n=1,2,3,\ldots$
        In the latter, our approach shows that the Fock state probability distribution of the initial state is conserved but the amplitudes suffer a phase shift proportional to $\pi/2$ that results in partial quantum state transfer for any other network size.
	\end{abstract}
	
	
	\maketitle
\section{Introduction}

Integrated photonic quantum technologies \cite{Grafe2020p073001,Wang2020p273} promise higher speed, lower energy loss or greater bandwidth, to mention a few, that may impact all optical quantum communications and computing.
Resonator networks are available to fulfill these promises in multiple platforms using evanescent field coupling between modes of individual high-Q resonators; for example, whispering gallery modes in optical resonators \cite{Gorodetsky1999p147,Yariv1999p711,Cai2020p105968}, photonic crystal cavities \cite{Notomi2008p741,Matsuda2014p2290,Cai2013p141118}, or femtosecond-laser-written waveguide arrays \cite{Itoh2006p620,Dreisow2008p2689,Szameit2010p163001}.

Here, we focus on the transfer of non-classical states of light required, for example, by switching and routing. 
We draw inspiration from the so-called $J_{x}$ photonic lattice where experimental perfect state transfer for non-entangled \cite{Bellec2012p4504,Perez2013p012309} and entangled \cite{Chapman2016p11339} photonic qubits was demonstrated using an array of eleven coupled waveguides.
Recent advances on reconfigurable nanoelectronical networks allowed perfect coherent transfer on-chip \cite{Tian2020p174303}.
On the theoretical side, there is a recent report of quantum state transfer of two-photon Fock and NOON states but not of squeezed and coherent states for twenty coupled waveguides \cite{Swain2020p035202}.
The latter left the question on the feasibility of nonclassical state transfer in such resonator networks open and we aim to answer it.

In the following, we deal with the idea of normal modes for a completely connected resonator network described by an auxiliary Hermitian coupling matrix feasible of unitary diagonalization, Sec. \ref{sec:S1}.
Then, in Sec. \ref{sec:S2}, we explore the idea of quantum state transfer in such coupled resonator network to show that it reduces to find the structure of an auxiliary evolution matrix  given in terms of orthogonal polynomials. 
In order to provide a particular example, we study the quantum analog of photonic lattice with an underlying $su(2)$ symmetry, that includes the so-called $J_{x}$ photonic lattice, and give its auxiliary unitary, diagonal and evolution matrices in terms of Gauss Hypergeometric function that allows us to discuss full and partial quantum state reconstructions and transfer in Sec. \ref{sec:S3}.
We close with our conclusions in Sec. \ref{sec:S4}.
	
\section{Normal modes of an resonator network} \label{sec:S1}

We follow a Schr\"odinger picture equivalent of a Heisenberg picture method proposed by one of us \cite{RodriguezLara2011p053845}.
Let us start from a completely connected resonator network, 
\begin{eqnarray}
	\frac{\hat{H}}{\hbar} = \sum_{j} \omega_{j} \hat{a}_{j}^{\dagger} \hat{a}_{j} + \sum_{j \ne k} \left( g_{j,k} \hat{a}_{j}^{\dagger} \hat{a}_{j+k} + g_{j,k}^{\ast} \hat{a}_{j+k}^{\dagger} \hat{a}_{j} \right),
\end{eqnarray}
and rewrite it in a simplified form, 
\begin{eqnarray}
	\frac{\hat{H}}{\hbar} = \sum_{j,k}  \left[\mathbf{M}\right]_{j,k}  \hat{a}_{j}^{\dagger}  \hat{a}_{k},
\end{eqnarray}
in terms of an auxiliary Hermitian coupling matrix $\mathbf{M}$ with diagonal elements provided by the resonator frequencies, $\left[\mathbf{M}\right]_{j,j} = \omega_{j}$, and off-diagonal elements by the coupling strengths, $\left[\mathbf{M}\right]_{j,k} = g_{j,k} = \left[\mathbf{M}\right]_{k,j}^{\ast}$.
In the following, we make use of the fact that all tri-diagonal auxiliary Hermitian coupling matrices are feasible of unitary diagonalization through orthogonal polynomials \cite{Bruschi2007p9793} as well as some penta-diagonal symmetric matrices \cite{Andelic2020p}.
On a case-by-case basis, this may extend to block tri- and penta-diagonal matrices.
In these and other cases, there exists a unitary matrix transformation $\mathbf{D}$ that diagonalizes our auxiliary Hermitian matrix, 
\begin{eqnarray}
	\mathbf{M} = \mathbf{D}^{\dagger} \mathbf{\Lambda} \mathbf{D}, 
\end{eqnarray}
with a real diagonal matrix $\mathbf{\Lambda}$. 
Here, it is possible to define normal modes,
\begin{eqnarray}
	\hat{c}_{p} = \sum_{j} \left[ \mathbf{D} \right]_{p,j} \hat{a}_{j}, 
\end{eqnarray}
that provide a diagonal representation of our Hamiltonian, 
\begin{eqnarray}
	\frac{\hat{H}}{\hbar} = \sum_{p} \lambda_{p} \hat{c}_{p}^{\dagger} \hat{c}_{p}, 
\end{eqnarray}	
where the normal mode frequencies are the elements of the diagonal matrix, $ \left[ \mathbf{\Lambda}\right]_{p,q} = \lambda_{p} \delta_{p,q}$.
The time evolution in terms of these normal modes,
\begin{eqnarray}
	\hat{U}(t) = e^{ - i \sum_{p} \lambda_{p} \hat{c}_{p}^{\dagger} \hat{c}_{p} t },
\end{eqnarray} 
may be complicated but straightforward to calculate as we will show in the following section.

\section{Quantum state transfer} \label{sec:S2}

For the sake of providing a practical example, let us focus on transferring an arbitrary initial state at the $m$-th resonator, 
\begin{eqnarray}
	\vert \psi (0) \rangle = \sum_{k=0}^{\infty} \alpha_{k} \vert k \rangle_{m}, \qquad \sum_{k=0}^{\infty} \vert \alpha_{k} \vert^2 = 1,
\end{eqnarray}
into the $n$-th resonator at some given transfer time $\tau > 0$. 
In order to discern the time evolution of the initial state, we may rewrite each Fock state in terms of creation operators, $ \vert k \rangle_{m} = (k!)^{-1/2} \hat{a}_{m}^{\dagger k} \vert 0 \rangle$, and expand the creation operator in terms of the normal modes, $ \hat{a}_{u}^{\dagger} = \sum_{p} \left[ \mathbf{D} \right]_{p,m} \hat{c}_{p}^{\dagger}$.
Then, the time evolution of the initial state,
\begin{eqnarray}
	\vert \psi (t) \rangle &=& \sum_{k=0}^{\infty} \alpha_{k} \, \frac{1}{\sqrt{k!}} \hat{U}(t) \left[  \sum_{p} \left[ \mathbf{D} \right]_{p,m} \hat{c}_{p}^{\dagger} \right]^{k} \vert 0 \rangle, \\
	&=& \sum_{k=0}^{\infty} \alpha_{k} \, \frac{1}{\sqrt{k!}}  \left[  \sum_{p} \left[ \mathbf{D} \right]_{p,m} e^{- i \lambda_{p} t} \hat{c}_{p}^{\dagger} \right]^{k}  \hat{U}(t) \vert 0 \rangle, \\
	&=& \sum_{k=0}^{\infty} \alpha_{k} \, \frac{1}{\sqrt{k!}}  \left[  \sum_{p,q} \left[ \mathbf{D} \right]_{p,m} e^{- i \lambda_{p} t} \left[\mathbf{D} \right]^{\ast}_{p,q} \hat{a}_{q}^{\dagger} \right]^{k}  \vert 0 \rangle,
\end{eqnarray} 
where we use the fact that the vacuum state is an eigenstate of the system and does not evolve, $\hat{U}(t) \vert 0 \rangle = \vert 0 \rangle$, and move back into the localized modes, $ \hat{c}_{p}^{\dagger} = \sum_{q} \left[ \mathbf{D} \right]^{\ast}_{p,q} \hat{a}_{q}^{\dagger}$.
In order to simplify our notation,
\begin{eqnarray}
	\vert \psi (t) \rangle = \sum_{k=0}^{\infty} \alpha_{k} \, \frac{1}{\sqrt{k!}} \left[ \sum_{q} \left[\mathbf{U}\right]_{m,q} \, \hat{a}_{q}^{\dagger} \right]^k \vert 0 \rangle,
\end{eqnarray}
we define an auxiliary evolution matrix, 
\begin{eqnarray}
	\mathbf{U}  = \mathbf{D}^{\dagger} e^{- i \mathbf{\Lambda} t} \mathbf{D},
\end{eqnarray}
provided by the unitary decomposition of the auxiliary Hermitian coupling matrix. 

At this point, our problem of transferring the localized initial state from the $m$-th to the $n$-th resonator reduces to find a unitary decomposition of the auxiliary Hermitian matrix, whenever it is possible, calculate the auxiliary evolution matrix elements, and find a transfer time $\tau$ where the auxiliary evolution matrix has the desired transfer matrix shape.
Thanks to the work on orthogonal polynomials for tri-diagonal Hermitian and penta-diagonal symmetric matrices, this process may simplify for networks whose effective Hermitian coupling matrix has these characteristics.
In the other hand, we may look at the inverse problem, starting from a desired transfer matrix shape, find an auxiliary Hermitian matrix that allows us to define a resonator network. 
The latter is beyond the scope of this contribution. 
In the following section, we look at the former using a well known quantum resonator network, and study its ability to produce quantum state transfer using our normal mode approach.

\section{SU(2) resonator network} \label{sec:S3}

In photonics, there exists a well-known photonic lattice with an underlying $su(2)$ symmetry that provides coherent transfer of classical light and single-photon states \cite{Perez2013p012309,Perez2013p022303}. 
In the quantum regime, it has been used to discuss the idea of synthetic dimensions \cite{Lustig2019p356}. 
We will work with a general form \cite{Villanueva2015p22836} that translates into a coupled resonator network with the following Hamiltonian, 
\begin{eqnarray}
	\frac{\hat{H}}{\hbar} = \sum_{m=0}^{2j} \omega_{m} \hat{a}_{m}^{\dagger} \hat{a}_{m} + \left( g_{m} \hat{a}_{m}^{\dagger} \hat{a}_{m+1} + g_{m+1} \hat{a}_{m+1}^{\dagger} \hat{a}_{m} \right),
\end{eqnarray}
where the annihilation (creation) operators for the localized modes in the $m$-th resonator are $\hat{a}_{j}$ ($\hat{a}_{j}^{\dagger}$) and we have $2j+1$ resonators in total.
The frequency of each resonator is $\omega_{m} = \omega_{0} + \omega \left( m - j  \right)$ and the nearest neighbours couplings is $ g_{m} = g \sqrt{m (2 j + 1 - m)} $.
The Hamiltonian conserves the total photon number operator, 
\begin{eqnarray}
	\hat{N} = \sum_{m=0}^{2j} \hat{a}_{m}^{\dagger} \hat{a}_{m},
\end{eqnarray}
that allows us to change into a reference frame, $\vert \psi \rangle = e^{-i \omega_{0} \hat{N} t } \vert \psi_{1} \rangle$, where the effective Hamiltonian, 
\begin{eqnarray}
	\frac{\hat{H}_{1}}{\hbar} = \sum_{m,n=0}^{2j} \left[\mathbf{M}\right]_{m,n}  \hat{a}_{m}^{\dagger} \hat{a}_{n} ,
\end{eqnarray}
is related to the auxiliary Hamiltonian coupling matrix,
\begin{eqnarray}
	\left[\mathbf{M}\right]_{m,n} = \omega \left( m - j  \right) \delta_{m,n} +  g  \left[ \sqrt{n (2 j + 1 - n)} \, \delta_{m,n+1} + \sqrt{m (2 j + 1 - m)} \, \delta_{m,n+1} \right],
\end{eqnarray}
with an underlying $su(2)$ symmetry with Bargmann parameter $j= n/2$ with $n=1,2,3, \ldots$ that connects with the $2j+1$ total number of resonators.
This auxiliary Hamiltonian matrix yields the normal modes, 
\begin{eqnarray}
	\frac{\hat{H}_{1}}{\hbar} = \sum_{m=0}^{2j} \mathbf{\Lambda}_{m,n}  \hat{c}_{m}^{\dagger} \hat{c}_{n}, \qquad  	\hat{c}_{m} = \sum_{n=0}^{2j} \left[\mathbf{D}\right]_{m,n} \hat{a}_{n}, 
\end{eqnarray}
in terms of the auxiliary unitary, diagonal, and evolution auxiliary matrices,
\begin{eqnarray}
	\left[\mathbf{D}\right]_{m,n} &=& f(m,n,A_{\pm},A_{z}) \\
	\left[ \mathbf{\Lambda} \right]_{m,n} &=&  \sqrt{ w^2 + 4 g^2} \, \left( m - j \right) \delta_{m,n}, \\
	\left[ \mathbf{U}(t) \right]_{m,n} &=& f\left[m,n,B_{\pm}(t), B_{z}(t)\right],
\end{eqnarray} 
in that order, related to the Wei-Norman decomposition parameters for the $su(2)$ algebra \cite{Ban1993p1347}, 
\begin{eqnarray}
	A_{\pm} = \pm \tan \theta, \quad \sqrt{A_{z}} = \sec \theta, \quad \tan 2\theta = \frac{2g}{\omega},
\end{eqnarray}
and
\begin{eqnarray}
	B_{\pm}(t) = - i \frac{2g \sin \frac{1}{2} \Omega t}{\Omega \cos \frac{1}{2} \Omega t + i \omega \sin \frac{1}{2} \Omega t} , \qquad
	\sqrt{B_{z}(t)} = \frac{\Omega}{\Omega \cos \frac{1}{2} \Omega t + i \omega \sin \frac{1}{2} \Omega t}, 
\end{eqnarray} 
where we express parameters $A_{z}$ and $B_{z}(t)$ in terms of their square root to highlight the importance of branch cuts when working with this class of resonator networks. 
We define an effective resonator network frequency,
\begin{eqnarray}
	\Omega = \sqrt{ w^2 + 4 g^2},
\end{eqnarray}
that will be at the core of our analysis.
In these expressions, we define the matrix elements,
\begin{eqnarray}
	f(m,n,X_{\pm},X_{z}) = \left\{ 
	\begin{array}{ll} 	
		\frac{1}{(n-m)!} \sqrt{\frac{n! (2j-m)!}{m! (2j-n)!}} X_{-}^{n-m} X_{z}^{m-j} \,_{2}\mathrm{F}_{1}\left[-m, 2j +1-m;n-m+1;-\frac{X_{+}X_{-}}{X_{z}} \right], 	& m<n, \\ 
		X_{z}^{m-j} \,_{2}\mathrm{F}_{1}\left[-m, 2j +1-m;1;-\frac{X_{+} X_{-}}{X_{z}} \right],	& m=n, \\
		\frac{1}{(m-n)!}  \sqrt{\frac{m! (2j-n)!}{n! (2j-m)!}} X_{+}^{m-n} X_{z}^{n-j} \,_{2}\mathrm{F}_{1}\left[-n, 2j +1-n;m-n+1;-\frac{X_{+} X_{-}}{X_{z}} \right],	& m>n,
	\end{array}  \right. \nonumber
\end{eqnarray}
in terms of the Gaussian Hypergeometric function $_{2}\mathrm{F}_{1}(x_{1},x_{2}; y; z)$ \cite{NIST:DLMF}.
Again, we must be cautious with the ramifications of the Wei-Norman decomposition parameter $X_{z}$.
For integer Bargmann parameters, $j=1,2,3\ldots$, we may use its absolute value, otherwise, $j= 1/2, 3/2, 5/2, \ldots$, we need to work with the definitions above.
While this analytic result may seem complicated, it provides great insight about quantum transport in the resonator network.

\subsection*{Quantum state reconstruction}

The Wei-Norman decomposition parameters, $B_{\pm}(t)$ and $B_{z}(t)$, of the auxiliary evolution matrix suggest an oscillatory behaviour with period, 
\begin{eqnarray}
	\tau_{r} = \frac{4 \pi}{\Omega} r, \qquad r=1,2,3,\ldots
\end{eqnarray}
that leads to decomposition parameters,
\begin{eqnarray}
	\lim_{t \rightarrow \tau_{r}} B_{\pm}(t) = 0 , \\
	\lim_{t \rightarrow \tau_{r}} \sqrt{B_{z}(t)} = 1, 
\end{eqnarray} 
pointing to a relevant Hypergeometric function,
\begin{eqnarray}
	 \lim_{t \rightarrow \tau_{r}}  \,_{2}\mathrm{F}_{1}\left[-m, 2j +1-m;1; -\frac{B_{p}(t) B_{m}(t)}{B_{z}(t)}  \right] = 1, 
\end{eqnarray}
that yields an auxiliary evolution matrix equal to the identity,
\begin{eqnarray}
	\lim_{t \rightarrow \tau_{r}} \mathbf{U}(t) = \mathbf{1}.
\end{eqnarray}
This result suggest that the initial quantum state,
\begin{eqnarray}
	\vert \psi(\tau_{r} ) \rangle =  \sum_{k=0}^{\infty} \alpha_{k} \vert k \rangle_{m},
\end{eqnarray}
will ideally reconstruct at this time.

\subsection*{Partial state reconstruction}

Something interesting happens at half the reconstruction time, 
\begin{eqnarray}
	\tau_{r/2} = \frac{2 \pi}{\Omega} r, \qquad r=1,3,5,\ldots
\end{eqnarray}
that leads to decomposition parameters,
\begin{eqnarray}
	\lim_{t \rightarrow \tau_{r/2}} B_{\pm}(t) &=& 0 , \\
	\lim_{t \rightarrow \tau_{r/2}} \sqrt{B_{z}(t)} &=& -1, 
\end{eqnarray} 
pointing to a relevant Hypergeometric function,
\begin{eqnarray}
	\lim_{t \rightarrow \tau_{r/2}}  \,_{2}\mathrm{F}_{1}\left[-m, 2j +1-m;1; -\frac{B_{p}(t) B_{m}(t)}{B_{z}(t)}  \right] = 1, 
\end{eqnarray}
that yields an auxiliary evolution matrix,
\begin{eqnarray}
	\lim_{t \rightarrow \tau_{r/2}} \mathbf{U}(t) = (-1)^{2j} \mathbf{1}.
\end{eqnarray}
equal to minus the identity for half odd integer, $j=1/2,3/2, 5/2, \dots$ and the identity for integer, $j=1,2,3,\ldots$ Bargmann parameters.
Thus, the initial quantum state will ideally reconstruct,
\begin{eqnarray}
	\vert \psi(\tau_{r/2} ) \rangle =  \sum_{k=0}^{\infty} \alpha_{k} \vert k \rangle_{m}, \qquad j=1,2,3, \ldots
\end{eqnarray}
only for odd sized networks. 
For networks of even size, 
\begin{eqnarray}
	\vert \psi(\tau_{r/2} ) \rangle =  \sum_{k=0}^{\infty} (-1)^{k} \alpha_{k} \vert k \rangle_{m}, \qquad j=1/2,3/2,5/2, \ldots
\end{eqnarray}
we recover the even components of the initial state and the odd ones will show a $\pi$-phase shift. 
In other words, the state will show the same probability distribution in terms of Fock states that the initial state, $P(n) = \vert \langle n \vert \psi(\tau_{r/2}) \rangle \vert ^2 = \vert \langle n \vert \psi(0) \rangle \vert ^2 = \vert \alpha_{n} \vert^2$, but the fidelity will be less than the unit, $\mathcal{F} = \vert \langle \psi(0) \vert \psi(\tau_{r/2}) \rangle \vert ^2 = \vert \sum_{k=0}^{\infty} (-1)^{k} \vert \alpha_{k} \vert^2 \vert^2 \le 1$ unless the initial state only has even or odd components.
For example, squeezed vacuum or even (odd) cat coherent states will fully reconstruct at this particular time but coherent states will not.

\subsection*{Perfect and partial quantum state transfer}

\begin{figure}
    \centering
    \includegraphics{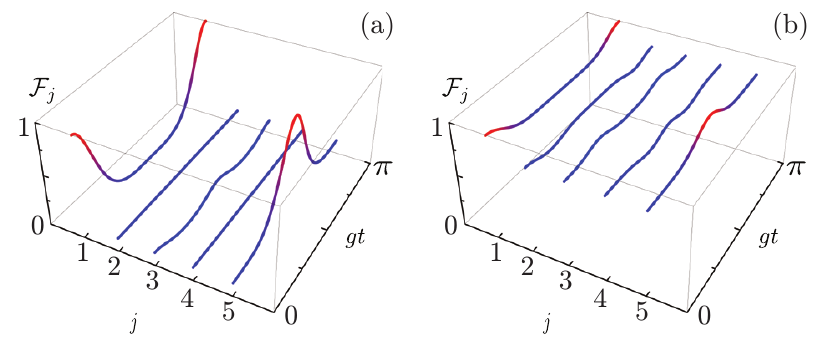}
    \caption{Fidelity in a so-called $J_{x}$ resonator network with five elements, $j = 2$, for an initial (a) single photon coherent state, $\alpha = 1$, and (b) squeezed vacuum state with squeezing amplitude $r = \sqrt{0.2}$. As expected for a network of this size, perfect quantum state transfer from the first to the last resonator occurs at $g t = \pi / 2$ and perfect quantum state reconstruction at the initial resonator occurs at $gt = \pi$ as witnessed by a unit value of the fidelity.}
    \label{fig:Fig1}
\end{figure}

In the classical and single-excitation quantum regime, the so-called $J_{x}$ oscillator network produces coherent quantum transfer. 
This array requires all resonators in the network to be identical.
In consequence, we deal with an effective nil frequency, $\omega \rightarrow 0$, that produces Wei-Norman decomposition parameters,
\begin{eqnarray}
	\lim_{\omega \rightarrow 0} B_{\pm}(t) &=& - i \tan g t, \\
	\lim_{\omega \rightarrow 0} \sqrt{B_{z}(t)} &=& \sec g t, \\
	\lim_{\omega \rightarrow 0} \frac{ B_{+}(t)  B_{-}(t)}{B_{z}(t)} &=& - \sin^2 g t, 
\end{eqnarray}  
that helps us explore a quarter of the reconstruction time,  
\begin{eqnarray}
	\tau_{t} = \lim_{\omega \rightarrow 0} \frac{\pi}{\Omega} = \frac{\pi}{2 g},
\end{eqnarray}
where the individual parameters above indeterminate but the matrix elements,
\begin{eqnarray}
	\lim_{ t \rightarrow \tau_{t} } \lim_{\omega \rightarrow 0} f\left[ m,n,B_{\pm}(t), B_{z}(t) \right] = \delta_{m,2j+1-m},
\end{eqnarray}
provide us with an auxiliary evolution matrix, 
\begin{eqnarray}
		\lim_{ t \rightarrow \tau_{t} } \lim_{\omega \rightarrow 0} \mathbf{U}(t) = (-i)^{2j} \mathbf{J}.
\end{eqnarray}
proportional to the backward identity matrix $\mathbf{J}$. 
This result suggest that the initial quantum state will ideally transfer from the $m$-th resonator into the $(2j+1-m)$-th resonator, 
\begin{eqnarray}
	\vert \psi(\tau_{t} ) \rangle =  \sum_{k=0}^{\infty} \alpha_{k} \vert k \rangle_{2j+1-m}, \qquad j=2,4,6,8,\ldots
\end{eqnarray}
for networks with size equal to $4d+1$ resonators with $d=1,2,3,\ldots$.
Otherwise, we will obtain partial state transfer, 
\begin{eqnarray}
	\vert \psi(\tau_{t} ) \rangle =  \sum_{k=0}^{\infty} (-i)^{2 j k} \alpha_{k} \vert k \rangle_{2j+1-m}, \qquad j= 1/2, 1, 3/2, 5/2, \ldots
\end{eqnarray}
with phase shifts that depend on the size of the lattice and the Fock state component. 
Again, the state will show the same probability distribution in terms of Fock states that the initial state, $P(n) = \vert \langle n \vert \psi(\tau_{t}) \rangle \vert ^2 = \vert \langle n \vert \psi(0) \rangle \vert ^2 = \vert \alpha_{n} \vert^2$, for any given resonator network size but we will have full quantum state transfer only for resonator networks with even Bargmann parameter as the Fidelity, $\mathcal{F} = \vert \langle \psi(0) \vert \psi(\tau_{t}) \rangle_{n} \vert ^2 = \vert \sum_{k=0}^{\infty} (-i)^{2 j k} \vert \alpha_{k} \vert^2 \vert^2 \le 1$, only becomes the unit in such case, $j = 2 p$ with $p=1,2,3,\ldots$, for any given initial state. 
Figure \ref{fig:Fig1}(a) shows the evolution of the Fidelity for a coherent state, 
\begin{align}
    \vert \alpha \rangle_{1} = e^{-\frac{\vert \alpha \alpha^2}{2}} \sum_{p} \frac{\alpha^{p}}{\sqrt{p!}} \vert p \rangle,
\end{align}
with a mean photon number equal to one, $\alpha = 1$, starting in the first waveguide, $j=1$, and Fig. \ref{fig:Fig1}(b) that for a squeezed vacuum state,
\begin{align}
    \vert \alpha \rangle_{1} = \frac{1}{\sqrt{\cosh r}} \sum_{p} \left( -e^{i \varphi} \tanh r \right)^{p} \frac{\sqrt{2p!}}{2^{p} p!} \vert 2 p \rangle_{1},
\end{align}
with squeezed parameter value $ \xi = r e^{i \varphi} = \sqrt{0.2}$. 
In both cases, the fidelity starts with a unit value in the first resonator and signals perfect quantum state transfer to the last resonator at the expected normalized time $g t = \pi / 2$. 
Then, we observe perfect quantum state reconstruction at the initial resonator at the normalized time $g t = \pi$, again, as expected from our analysis.
The high fidelity baseline for the squeezed vacuum state is due to its high vacuum component for the small value of the squeeze parameter $r = \sqrt{0.2}$.
We compared our analytic predictions with a full brute force numerical propagator to good agreement. 
Our numerical propagator considers an approximate Hilbert space from zero to up to five excitation for each of the five resonators. 
This allows us to cover most of the information from the single photon coherent and $r = \sqrt{0.2}$ squeezed vacuum states; their norms in these subspaces are $\langle \alpha \vert \alpha \rangle = 0.99996$ and $\langle r e^{i \varphi} \vert r e^{i \varphi} \rangle = 0.99084$ that allow us to assume a good approximation.

\section{Conclusion} \label{sec:S4}

We showed that a normal modes approach provides a tractable framework for propagation of non-classical light in networks of coupled resonators.
Our approach yields propagation in terms of orthogonal polynomials that provide further insight into the dynamics in the network whenever it is possible to use a unitary diagonalization of the auxiliary Hermitian coupling matrix of the system; for example, resonator networks described by auxiliary tri-diagonal Hermitian coupling matrices and some penta-diagonal real symmetric coupling matrices. 

In particular, we studied a network with underlying $su(2)$ symmetry and were able to identify that it provides full quantum state reconstruction; that is, state transfer to the same initial site.
In addition, we showed that it offers both partial and full reconstruction at half the full reconstruction time.
In the case of partial reconstruction, the Fock state probability distribution is recovered but the amplitudes for odd Fock state components show a $\pi$-phase-shift with respect to the original.
Naturally, quantum states with only even (odd) components are fully reconstructed (up to an overall $\pi$-phase shift). 

We also explored quantum state transfer in the equivalent of the so-called $J_{x}$ photonic lattice. 
We showed that full quantum state transfer occurs at a quarter of the reconstruction time for coupled resonator networks with even Bargmann parameter; that is, networks composed by $ N = 4 n + 1$ resonators with $n=1,2,3, \ldots$.
The transfer occurs between the $m$-th and the $(N-m+1)$-th waveguides with $m=1,2,3,\ldots,N$. 
Otherwise, the Fock state probability distribution at the transfer site is identical to the original one but the amplitudes for Fock state components shows a phase shift proportional to an integer multiple of $\pi/2$. 

Our normal mode approach makes it straightforward to calculate correlations and other quantum quantities of interest in terms of orthogonal polynomials whenever a unitary diagonalization for the auxiliary Hermitian matrix exists.

\begin{acknowledgments}
	B.M.R.-L. acknowledges financial support and the hospitality of R.-K.L. at National Tsing-Hua University and is thankful to Benjamin Jaramillo \'Avila for his help formatting the figure.
\end{acknowledgments}
	

%

\end{document}